\documentclass[aps,prb,preprint]{revtex4}
\usepackage{psfig}
\begin{document}

\title{Quantum billiards and constrained random wave correlations}
\author{W. E. Bies and N. Lepore
%\thanks{bies@physics.harvard.edu}
\\ Department of Physics, Harvard University, \\
Cambridge, MA 02138\\
\vskip 0.1in 
E. J. Heller
%\thanks{heller@physics.harvard.edu}
 \\ Department of Physics and Department of
Chemistry, \\Harvard University, Cambridge,  MA
02138} 

\date{November 22, 2002}

\begin{abstract} 

We study chaotic eigenfunctions in wedge-shaped and rectangular regions
using a generalization of Berry's conjecture. An expression for the
two-point correlation function is derived and verified numerically.


\end{abstract}

\maketitle  

\section{Introduction}

According to Berry's conjecture,\cite{berry1} chaotic eigenfunctions behave 
locally like random superpositions of plane waves with wavevector $k$, where 
$k=(1/\hbar)\sqrt{E-V}$. This description is consistent with random matrix 
theory. The Gaussian random wave model does not account for the  
localization properties of eigenfunctions, 
such as scarring \cite{scar} and weak quantum ergodicity.\cite{weak-ergod}
In particular, 
for quantum billiards with Dirichlet boundary conditions, 
the wavefunction must vanish at the boundary, and thus no longer looks 
random in its vicinity. For straight 
boundaries, the problem can be solved by reducing the set of available plane 
waves to those which are antisymmetric with respect to reflection across the 
wall, thereby ensuring a zero value along that boundary; see 
Berry.\cite{berry2} 
If, for instance the boundary is given by the line $y=0$, the chaotic wave 
function will be composed of a seemingly random combination of plane waves of 
the form ${\rm sin} (k_yy) {\rm cos} (k_xx+\phi)$ where $k_x^2+k_y^2=k^2$ and 
$\phi$ is a random phase shift. Bies and Heller \cite{BH} discuss similar 
boundary effects in soft potentials.

Some criteria for unconfined random waves appear to be well satisfied
by eigenfunctions of chaotic billiards. McDonald and Kaufman
\cite{mcdonald-kaufman} and McDonald \cite{mcdonald} checked
for random nodal patterns, Gaussian statistics and Bessel function
correlations. In their original work on the stadium, 
these authors emphasized the qualitative appearance of nodal lines in the
eigenfunctions. The nodal lines appear to wander
randomly throughout the billiard, which is indicative of an isotropic
distribution of local wave vectors.
Furthermore, they found the wavefunction statistics to be Gaussian. 
However, the
correlation function is not a Bessel function for a general billiard.
 
We first generalize Berry's boundary-adapted form of the Gaussian
random wave model to wedge-shaped regions. The case of a 90$^\circ$
angle is immediately solved: 
one just antisymmetrizes with respect to reflections in both $x$ and $y$. The 
simplest non-trivial case consists of a wedge with an opening angle of 
60$^\circ$. 
We focussed on this example in what follows, although our method
extends to all 
opening angles of $\pi/n$ radians, where $n$ is a positive integer. We first 
determine the right boundary-adapted plane wave basis. The two-point 
correlation 
function can be used to compare our model to numerically generated 
chaotic eigenstates. For Gaussian random waves in free space, 
the later is known to give
\begin{equation}
\langle \psi({\bf x}) \bar{\psi}({\bf x}') \rangle = J_0( {\bf |x-x'|})
\end{equation}
where the average 
is taken over a set of eigenstates. As we shall see, 
the two point correlation function for 60$^\circ$ wedge
again reduces to a sum of Bessel functions. Our numerically
generated ensemble of wavefunctions is compared to this 
form of the correlation function, 
including the interference effects of two or more Bessel functions.

Our  second type of billiard consists of two semi-infinite parallel lines 
at $ x > 0, y = \pm a/2$, which are connected by a perpendicular wall
at $x = 0$. The wavefunction can be made to vanish on the back wall by 
antisymmetrizing it with respect to the y-axis. The parallel walls are not so 
readily handled. The antisymmetrization procedure extends over an infinite 
periodic array of such lines a distance $a$ apart. 
The two-point correlation function 
will be composed of an infinite sum of Bessel functions. Since the later 
diminsh 
rapidly with distance, only the nearest neighbours contribute, allowing us to 
compare our formula with its numerically determined value.

\section{The two-point correlation function}

\subsection{The wedge}

We first look for the properly antisymmetrized wavefunctions. Let 
\begin{equation}
\psi({\bf x})=\int d\theta a_\theta e^{-i {\bf k}_\theta \cdot {\bf x} 
+ i \delta_\theta}
\end{equation}
denote a random sum of plane waves. Here ${\bf k}_\theta=k{\rm cos}\theta
\hat{\bf x} + k {\rm sin}\theta \hat{\bf y}$, the $a_\theta$ are independent
Gaussian distributed random variables and the $\delta_\theta$ are independent
uniformly distributed random phase shifts. Let $\tilde{\psi}({\bf x})$ denote 
the desired boundary-adapted version of $\psi({\bf x})$. Since 
$\tilde{\psi}({\bf x})$ vanishes on 
the boundary of the wedge, it can be extended to a fictitious wavefunction 
living in the whole plane
by reflecting antisymmetrically across either boundary. 
To obtain $\tilde{\psi}({\bf x})$, we let $R_1$ denote reflections
with respect to one of the lines bounding the wedge and $R_2$ reflections
with respect to the other line. We would like to antisymmetrize
$\psi({\bf x})$ with respect to both of these reflections. The projection
operator $(1/4)(1-R_1)(1-R_2)$ does not work, however, because $R_1$ 
and $R_2$ do not commute. Instead we must resort to the following
group-theoretical construction. Let $R_3$ denote the reflection with
respect to the line that passes through the vertex of the wedge and
meets the two edges of wedge at a 60$^\circ$ angle. The product $R_1R_2$ 
of the two reflections $R_1$ and $R_2$ is a rotation through 120$^\circ$
centered on the vertex, and $R_2R_1$ is the opposite rotation. Thus, the set 
${1, R_1, R_2, R_3, R_1R_2, R_2R_1}$ forms a representation of the dihedral 
group $C_3$. It is now easy to see that $\tilde{\psi}({\bf x})$ transforms 
under the one-dimensional representation of $C_3$ that assigns a character
of -1 to the reflections and 1 to the rotations. Put another way, the
problem of going from $\psi({\bf x})$ to $\tilde{\psi}({\bf x})$ is
that of projecting onto this irreducible representation of $C_3$. Formally, 
this transformation is expressed as:\cite{tinkham}
\begin{equation}
\tilde{\psi}({\bf x}) = {1 \over \sqrt{6}} \sum_{A \in C_3} \chi(A) A \psi({\bf x}).
\end{equation}
Here, $A$ denotes the group elements 
$1,R_1,R_2,R_3,R_1R_2$ and $R_2R_1$, and the character
is given by $\chi(R_1)=\chi(R_2)=\chi(R_3)=-1$ and $\chi(1)=\chi(R_1R_2)
=\chi(R_2R_1)=1$. 
The normalization factor of $1/\sqrt{6}$ is chosen
to ensure that the 
two-point correlation function gives the free-space result far away from the 
wall.
If we then restrict $\tilde{\psi}({\bf x})$ to the wedge-shaped region,
we obtain the required boundary-adapted sum of plane waves.

Now that we have an explicit expression for $\tilde{\psi}({\bf x})$
we can use it to calculate the two-point correlation function
$C({\bf x},{\bf x+r})=\langle \tilde{\psi}({\bf x})
\bar{\tilde{\psi}}({\bf x+r})
\rangle$. The correlation function is a sum of thirty-six terms of the
form 
\begin{eqnarray}
\langle \chi(A) A \psi({\bf x}) \chi(B) B \bar{\psi}({\bf x+r}) \rangle &=&
\chi(A)\chi(B) \int d\theta e^{-i{\bf k}_\theta \cdot (A{\bf x}-B{\bf x})
+i{\bf k}_\theta \cdot B{\bf r}} \\
&=& \chi(A) \chi(B) \int d\theta e^{ik\rho{\rm cos}(\theta-\theta_0)} \\
&=& \chi(A)\chi(B)  J_0(k\rho)
\end{eqnarray}
Here $\rho=|A{\bf x}-B({\bf x+r})|$, that is,  
the distance from ${\bf x+r}$ to $B^{-1}A{\bf x}$.
The correlation function becomes
\begin{eqnarray}
\label{wedge_correlation}
C({\bf x},{\bf x+r}) &=& {1 \over 6} \sum_{A,B \in C_3} \chi(A)\chi(B)
J_0(k\rho_{B^{-1}A}) \\
&=& \sum_{C \in C_3} \chi(C)J_0(k\rho_{C^{-1}}) \\
&=& J_0(kr)-J_0(k\rho_{R_1})-J_0(k\rho_{R_2})
-J_0(k\rho_{R_3})+J_0(k\rho_{R_1R_2})+J_0(k\rho_{R_2R_1})
\end{eqnarray}
In the second line, we replaced $B$ by $AC$. The sum over $A$ then
becomes trivial, since $\chi(A)\chi(B)=\chi(A)\chi(A)\chi(C) = \chi(C)$.
In the last line $\rho_A$ denotes the distance from ${\bf x+r}$ to
$A{\bf x}$. 

This formula predicts that the two-point correlation function should
display the interference of Bessel functions centered at all the points
given by applying the group transformations on ${\bf x}$. 
Far from the boundary, 
Eq~\ref{wedge_correlation} reduces to the single Bessel function $J_0(kr)$, 
the value that it would have if no boundary were present. However, the 
interference is pronounced near the boundary, and more especially near
the vertex of the wedge.

\subsection{the rectangular corridor}

In the case of the rectangular corridor, the back wall and the semi-infinite 
walls need to be handled separately. The wavefunction at ${\bf x}$ in the 
$x<0$ half-plane is obtained by reflecting about the y-axis.
To extend the wavefunction in the $y$ direction, we reflect it 
with respect to the wall at $y=a/2$. By repeating this process for all 
$y = (1+2m)a/2$ 
where $m = \{..., -1, 0, 1, ...\}$, the whole plane is tiled with positive or
negative copies of $\psi({\bf x})$. 

Mathematically, the process by which $\tilde{\psi}({\bf x})$ is obtained is as 
follows. The initial billiard is placed on an infinitely long cylinder 
of circumference 
$2an$. Our initial x-axis is now parallel to the axis of the cylinder, whereas
the y-axis is wrapped around the cylinder and truncated at 
$y = (1 \pm 2n)a/2$. 
The wavefunction is first reflected across the y-axis.
Therefore the antisymmetrization in the x-direction transforms as a 
one-dimensional
representation of the reflection group ${\bf Z_2}$.
Rotations by multiples of $2a$ should give back the wavefunction, 
while reflections about any of the lines parallel to the x-axis
give its negative version. We thus obtain a 
representation of the dihedral group $C_n$. 
It is now clear that on the cylinder, $\tilde{\psi}({\bf x})$ transforms under
the one-dimensional representation of $C_n \otimes{\bf Z_2}$
that assigns a character of $-1$ for reflections and 1 for rotations. 
This time the transformation is given by as
\begin{equation}
\tilde{\psi}_n({\bf x}) = {1 \over \sqrt{4n}} \sum_{A \in C_n \otimes{\bf Z_2} } 
\chi(A) A \psi({\bf x}).
\end{equation}
As in the case of the wedge, 
the normalization factor is found by requiring that
the correlation function give $J_0(k\rho)$ far away from the wall.

The correlation function is given by:
\begin{eqnarray}
\label{corridor_correlation}
C({\bf x},{\bf x+r}) &=& {\rm lim}_{n \rightarrow \infty} {1 \over 4n} \sum_{A,B \in C_n \otimes{\bf Z_2}} \chi(A)\chi(B)
J_0(k\rho_{B^{-1}A}) \\
&=& {\rm lim}_{n \rightarrow \infty}  \sum_{C \in C_n \otimes{\bf Z_2}} \chi(C)J_0(k\rho_{C^{-1}}).
\end{eqnarray}
However, $J_0(k\rho_{C^{-1}}) \rightarrow 1/\sqrt{k\rho_{C^{-1}}}$. 
Hence, only the Bessel functions centered at nearby points
contribute, 
and the correlation function remains finite as we take the 
$n \rightarrow \infty$ limit. 

\section{Numerical results}

Here we apply the theory to check further the properties of random
waves in billiard systems. There are some obvious and some more subtle
modifications of random wave behavior known in closed chaotic
billiards. The wavefunction must vanish on  the boundary, and scarring 
affects some states in a nonrandom way.\cite{kaplan} 
The theory given here suggests new correlations which ought to be checked in chaotic systems.  
As we incorporate more about a specific billiard geometry into our
correlation functions, we are probing the properties of waves which
are ``as random as possible'' within the constraints, such as a wedge 
boundary being present. In checking the numerics of a billiard such as
the cone or stadium, we are really asking whether the eigenstates are 
truly random but for the constraint of interest, say a wedge boundary.  
Naturally this cannot be strictly true, since there are more
constraints that we haven't included. At the end of this strategy
comes a tautology: if, in a closed billiard we incorporate {\em all} 
the constraints, we have only the eigenstates left, as the {\em only} 
waves which are consistent with all the constraints!

\subsection{The wedge}
To validate our expressions for $\tilde{\psi}({\bf x})$ and 
$C({\bf x},{\bf x+r})$, we generated an ensemble of 500 eigenstates
near $k=200$ for a $60^\circ$ wedge which has been closed off by
a semicircle; see Fig~1. The eigenstates were found using the 
Boundary Integral Method. Our cone-shaped billiard has a circle diameter 
equal to 1, which is about 32 wavelengths across.
The Poincare section indicates that 
this billiard is chaotic. 
In Fig.~2, we show $C({\bf x},{\bf x+r})$ for (a) {\bf x}
on the symmetry line through the center of the billiard, and (b) 
for ${\bf x}$ placed within a wavelength of one of the edges. 
Both Fig.~2(a) and 2(b) display the expected
interference, but the later is especially pronounced 
in Fig.~2(b). Note that since ${\bf x}$ is on the symmetry line, 
only even states contribute to Fig. 2(a).

For comparison, Fig.~3 displays the theoretical predictions for
the two-point correlation functions of Fig.~2. Both plots are in good 
agreement with their numerical equivalent. In particular, 
the interference pattern of Fig.~2b is reproduced in Fig.~3b. However, 
the angular nodes are much more pronounced in the theoretical plots,
suggesting that an unaccounted-for
smoothing process is at work in the numerical experiment. 
We compared the error quantitatively using 
\begin{equation}
\label{error}
\frac{\int d^2r |C_{\rm num}({\bf x},{\bf x+r})-C_{\rm th}({\bf x},{\bf x+r})|^2}
{\int d^2r C_{\rm th}({\bf x},{\bf x+r})^2}.
\end{equation}
We obtained 0.52 for 
Fig.~2a and 0.15 for Fig.~2b. The discrepancy in the predicted and
experimental values is expected due to the difference in the angular nodes, 
and appears to be more pronounced far away from the billiard walls.

In Fig.~4, we take the angular average of the correlation functions of Fig.~2.
As was found by Li and Robnik,\cite{li-robnik}
Fig.~4a resembles the Bessel function $J_0(kr)$.
This should be the case whenever boundary effects contribute equally from
all sides. Fig.~4b also approximates a Bessel function because, as can be
seen from Fig.~2b, the boundary effects affect only a small range of angles
and furthermore are independent of radius.

\subsection{The rectangular corridor}

Once again, we took an ensemble of 500 eigenstates
near $k=200$, this time for a quarter stadium with a circular radius 
of $R=0.6$ and straight length $l=1.2$ (about 20 wavelengths
wide); see Fig.~5. The numerical correlation function is shown in Fig.~6 
for ${\bf x}$
(a) on the symmetry line through the center of the billiard, 
(b) within 1 $\lambda$ of the top wall,
(c) within 1 $\lambda$ of a corner
and (d) within 1 $\lambda$ of the back wall.
As expected, only the Bessel functions which are less than a few
wavelengths away contribute to  $C({\bf x},{\bf x+r})$.

The numerical correlation function was compared to the 
theoretical prediction using Eq.~\ref{error}. Here only the 
first few terms were kept in $C_{\rm th}({\bf x},{\bf x+r})$.
We obtain 0.55 for Fig.~6a, 0.23 for Fig.~6b, 0.03 for Fig.~6c
and 0.24 for Fig.~6d,
in agreement with our results for the cone billiard.

\section{Conclusion}

We have successfully extended the boundary-adapted Gaussian wave model
to a wedge-shaped region with opening angle 60$^\circ$. Our technique is 
readily generalized to any opening angle of the form $\pi/n$ for $n$ integer; 
one merely replaces the group elements of $C_3$ with those of the dihedral 
group $C_n$. We also solved the case of the semi-infinite rectangular corridor.
It is our hope that these result will stimulate more work on  
boundary effects in arbitrarily shaped billiards.

\section{Acknowledgments}

We would like to thank Doron Cohen for making available his Boundary
Integral Method code. This work was supported by the National Science
Foundation grant CHE-0073544.

\newpage

\begin{figure}
\centerline{
\psfig{file=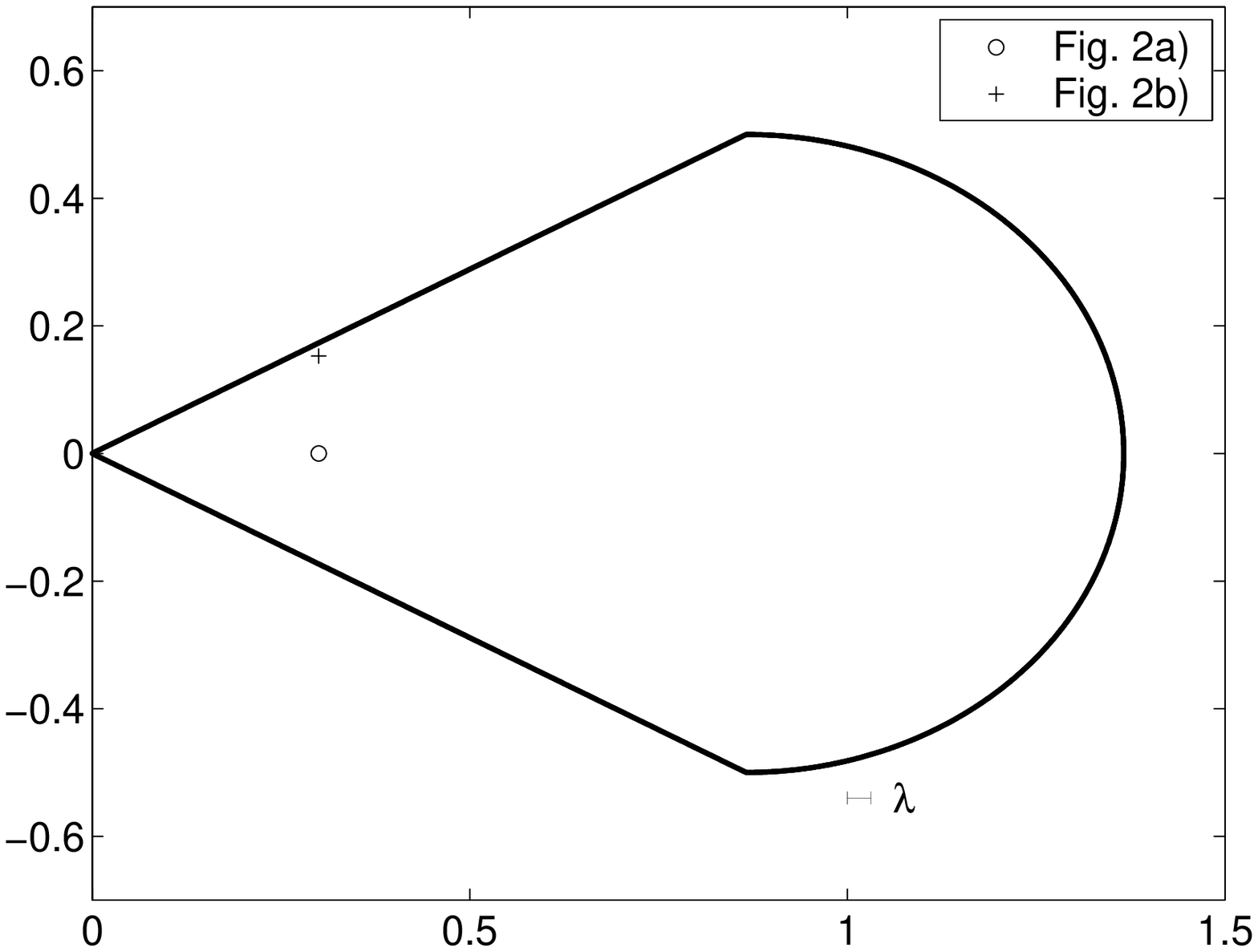,width=3in,bbllx=57pt,bblly=206pt,bburx=552pt,bbury=583pt,clip=}}
\vspace{0.1in}
\noindent
{\footnotesize {\bf FIG. 1.}  The cone billiard.}
\label{fig_1}
\end{figure}

\begin{figure}
\centerline{
\psfig{file=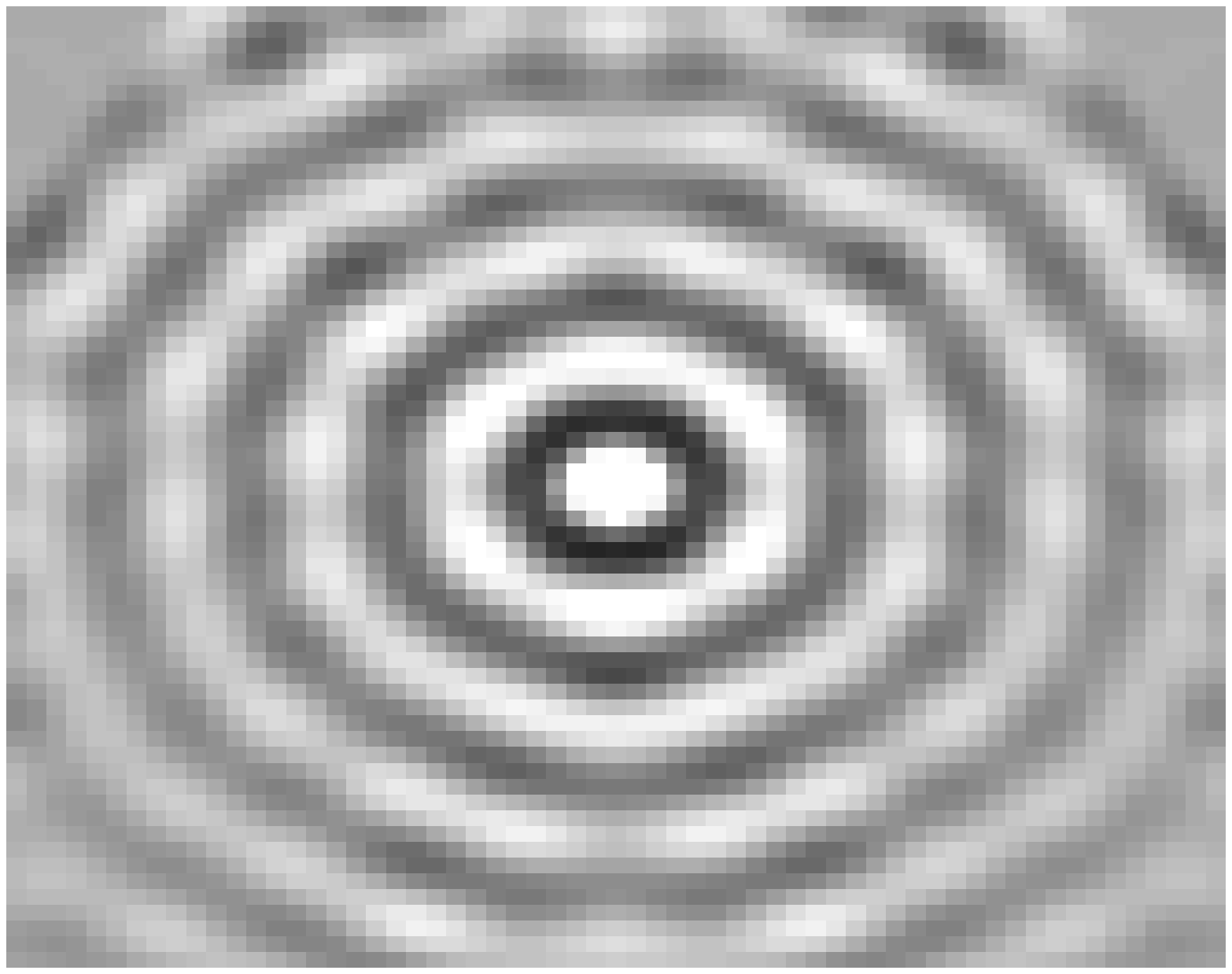,width=3in,bbllx=91pt,bblly=226pt,bburx=539pt,bbury=580pt,clip=}}
\vspace{0.1in}
\noindent
{\footnotesize {\bf FIG. 2(a).} Experimental correlation function for 
  ${\bf x} =
  (0.3, 0)$, on the symmetry axis through the center of the cone. 
  The grid is a square of side 0.129, \
  or about 8 wavelengths, 
centered on ${\bf x}$.}
\label{fig_2a}
\end{figure}

\begin{figure}
\centerline{
\psfig{file=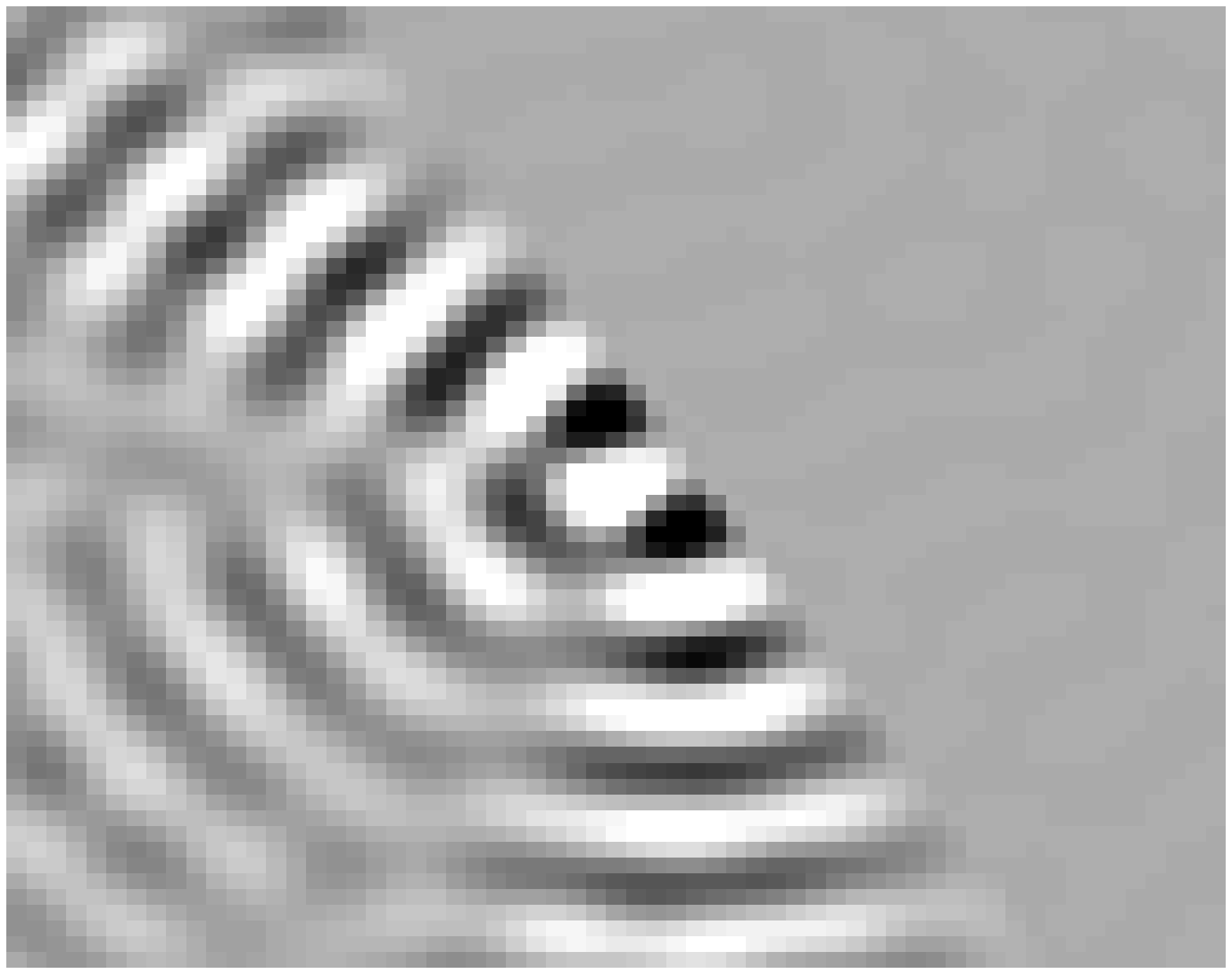,width=3in,bbllx=77pt,bblly=214pt,bburx=541pt,bbury=583pt,clip=}}
\vspace{0.1in}
\noindent
{\footnotesize {\bf FIG. 2(b).} Same as Fig.~2(a) for ${\bf x} = (0.3, 0.153)$,
near the upper straight edge of the cone.}
\label{fig_2b}
\end{figure}

\begin{figure}
\centerline{
\psfig{file=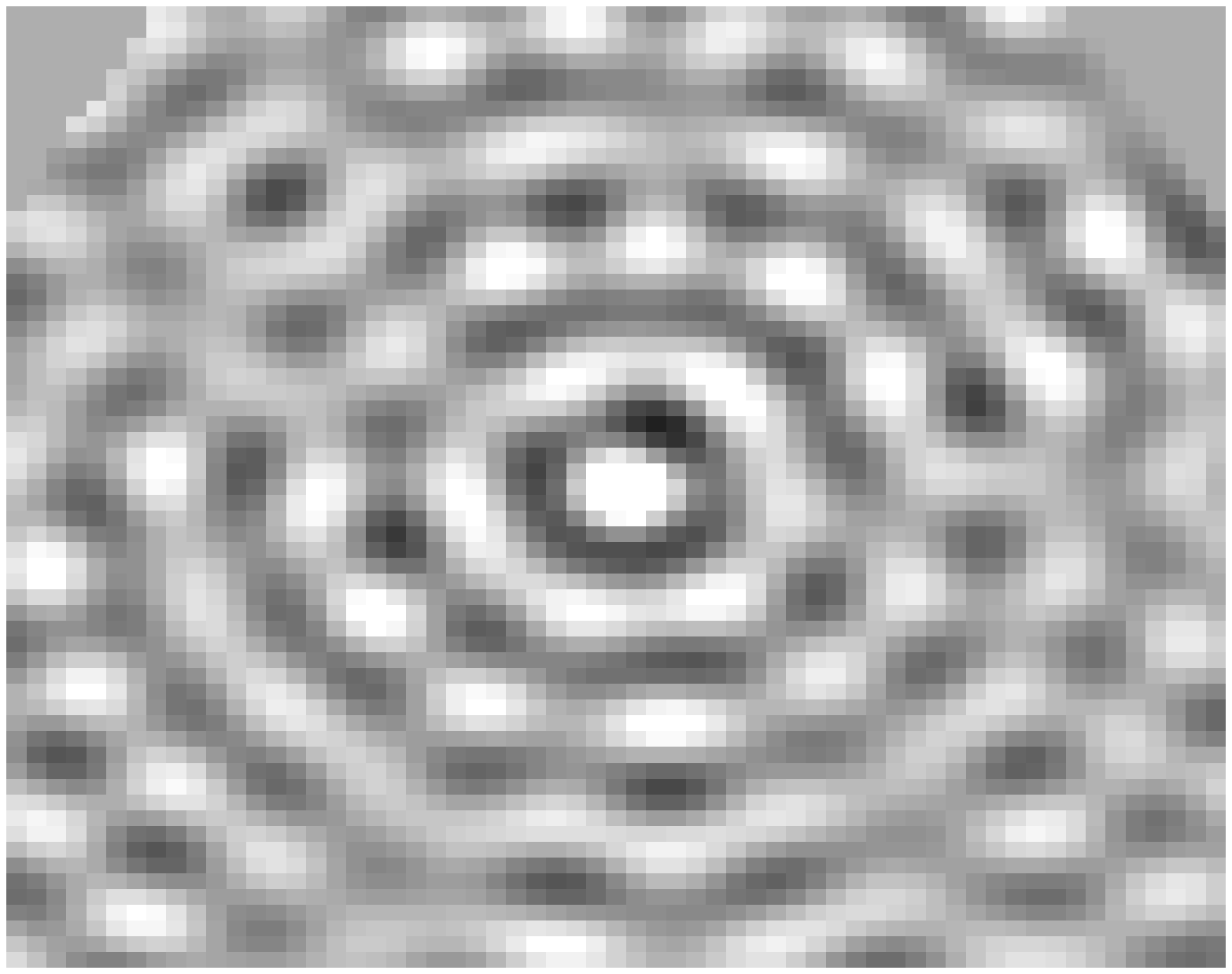,width=3in,bbllx=77pt,bblly=214pt,bburx=540pt,bbury=580pt,clip=}}
\vspace{0.1in}
\noindent
{\footnotesize {\bf FIG. 3(a).} Theoretical 
  correlation function for ${\bf x} = (0.3, 0)$ (same grid as in Fig.~2(a)).}
\label{fig_3a}
\end{figure}

\begin{figure}
\centerline{
\psfig{file=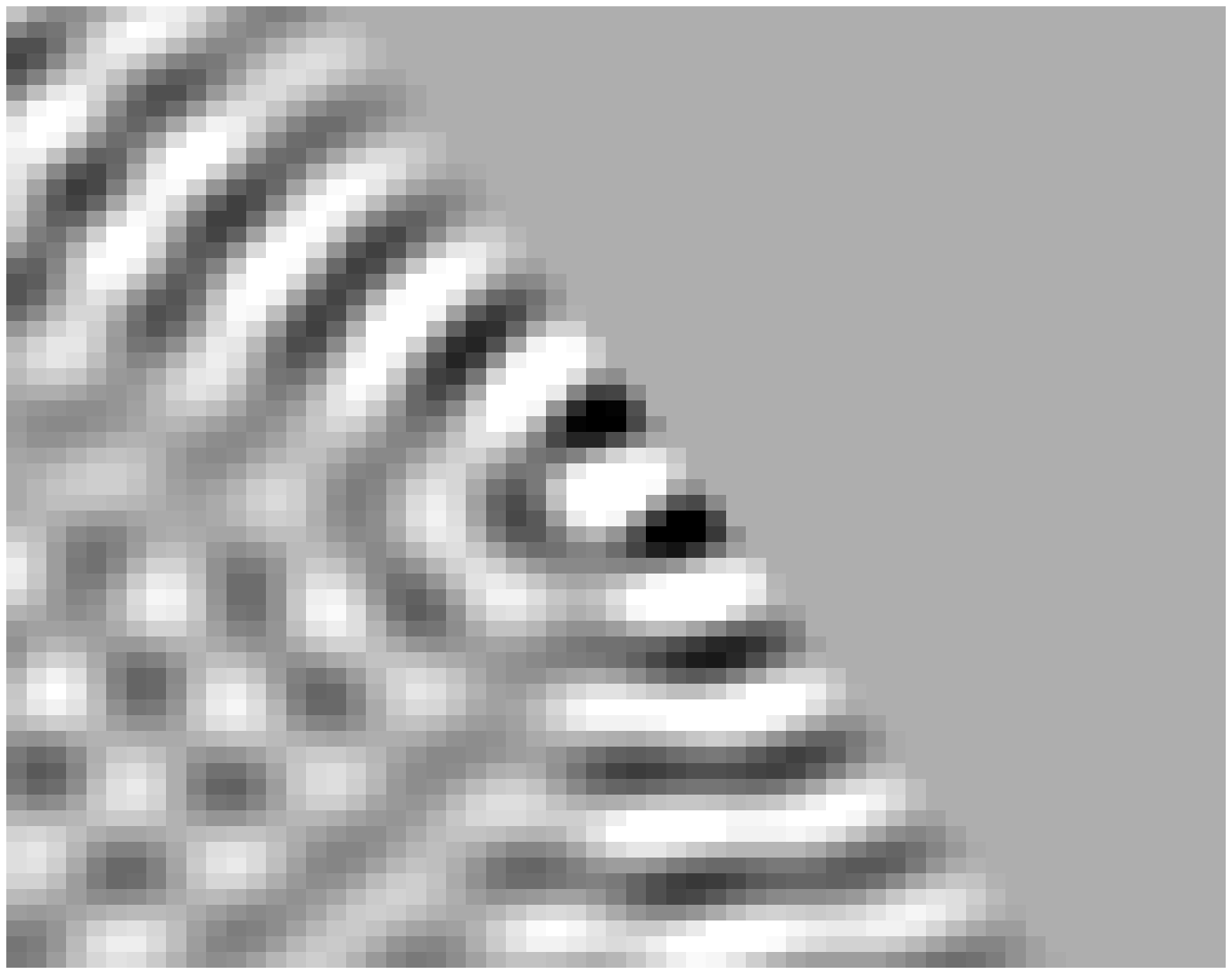,width=3in,bbllx=77pt,bblly=213pt,bburx=540pt,bbury=582pt,clip=}}
\vspace{0.1in}
\noindent
{\footnotesize {\bf FIG. 3(b).} Same as Fig.~3(a) for ${\bf x} = (0.3, 0.153)$
(same grid as in Fig.~2(b)).}
\label{fig_3b}
\end{figure}

\begin{figure}
\centerline{
\psfig{file=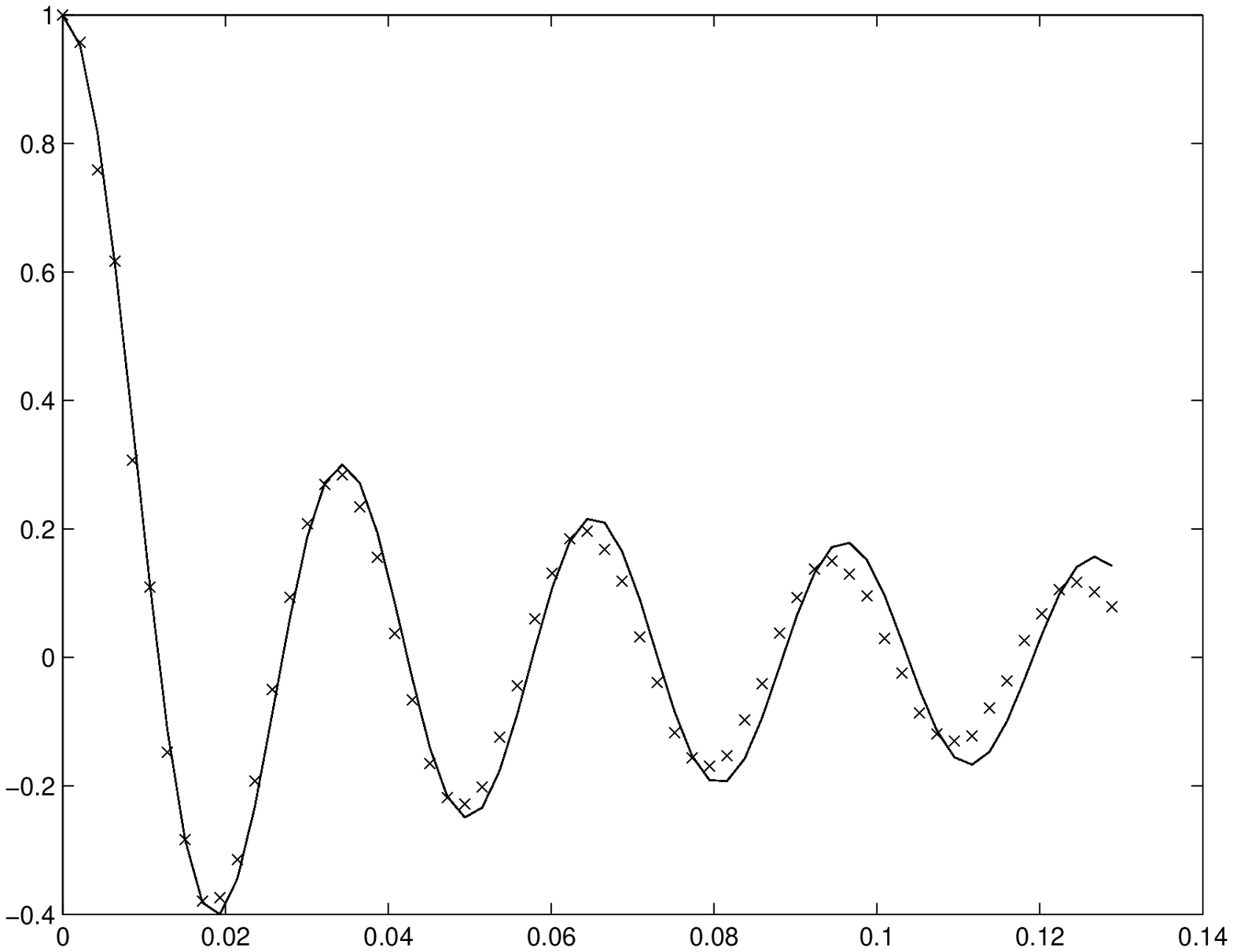,width=3in,bbllx=57pt,bblly=205pt,bburx=550pt,bbury=587pt,clip=}}
\vspace{0.1in}
\noindent
{\footnotesize {\bf FIG. 4(a).} Angular average of 
  $C({\bf x+r},{\bf x})$ vs $r$ for ${\bf x} = (0.3, 0)$ (same grid as in 
  Fig.~2(a)). 
  Solid line, Bessel function $J_0(kr)$; $\times$, numerical data.
}
\label{fig_4a}
\end{figure}

\begin{figure}
\centerline{
\vspace{0.1in}
\noindent
\psfig{file=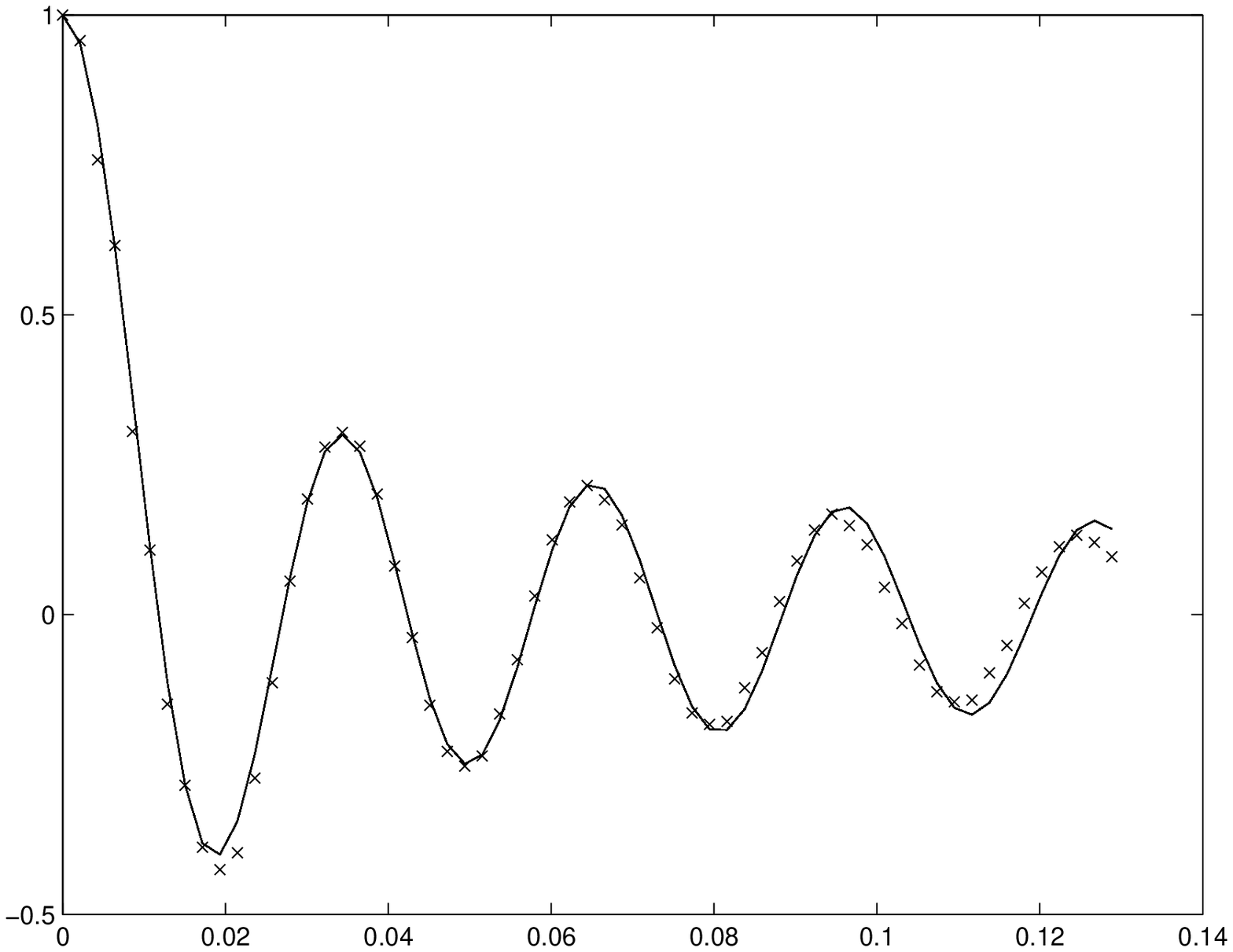,width=3in,bbllx=57pt,bblly=205pt,bburx=550pt,bbury=587pt,clip=}}
{\footnotesize {\bf FIG. 4(b).} Same as Fig.~4(a) for ${\bf x} = (0.3,
  0.153)$ (same grid as in Fig.~2(b)).}
\label{fig_4b}
\end{figure}

\begin{figure}
\centerline{
\psfig{file=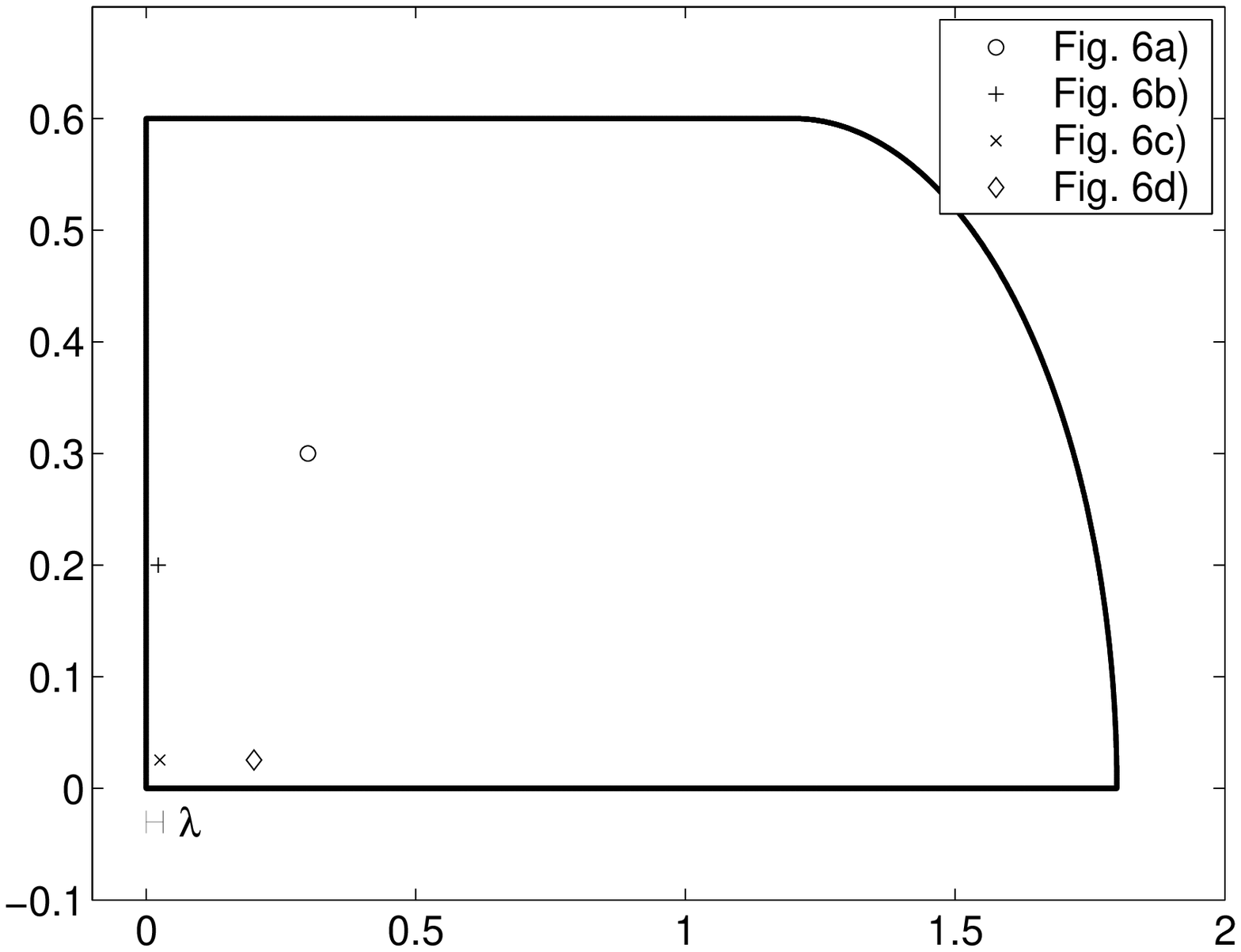,width=3in,bbllx=57pt,bblly=206pt,bburx=545pt,bbury=591pt,clip=}}
\vspace{0.1in}
\noindent
{\footnotesize {\bf FIG. 5.}  The stadium billiard.}
\label{fig_5}
\end{figure}

\begin{figure}
\centerline{
\psfig{file=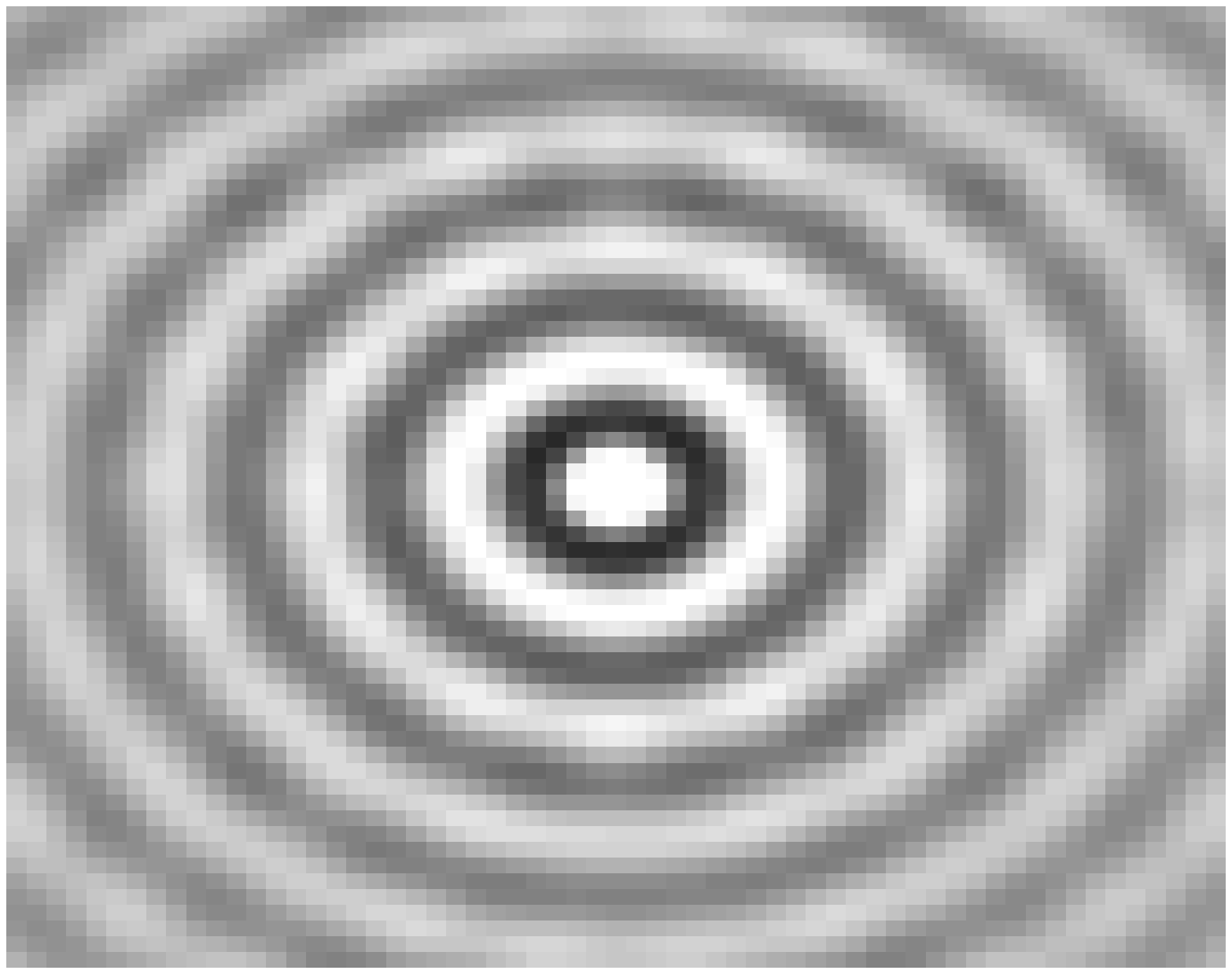,width=3in,bbllx=77pt,bblly=213pt,bburx=540pt,bbury=583pt,clip=}}
\vspace{0.1in}
\noindent
{\footnotesize {\bf FIG. 6(a).} 
  Experimental correlation function for ${\bf x} = (0.3, 0.3)$, the mid-point
  of the square part of the stadium. The
  grid is a square of side 0.140, or about 8 wavelengths, centered on
  ${\bf x}$. }
\label{fig_6a}
\end{figure}

\begin{figure}
\centerline{
\psfig{file=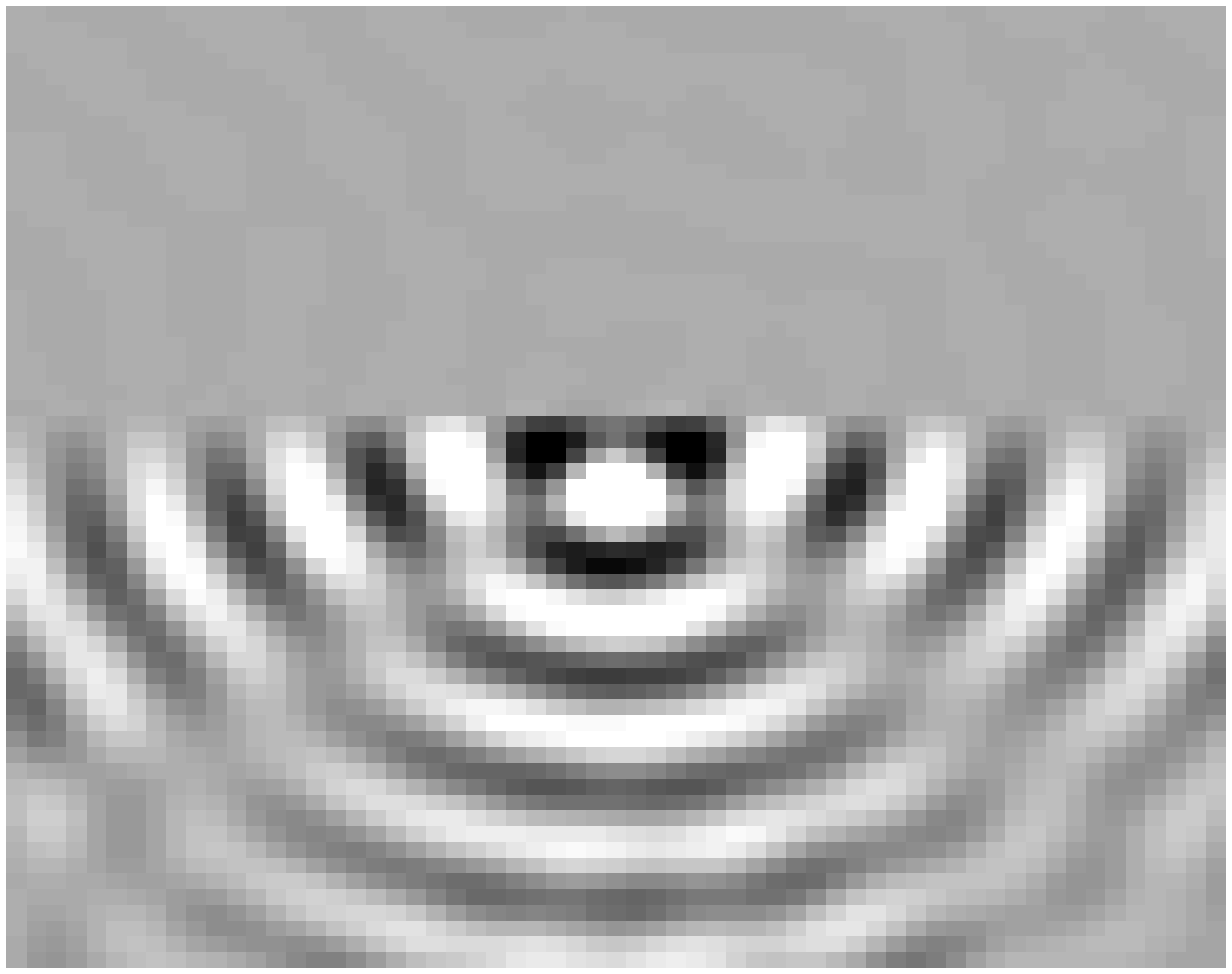,width=3in,bbllx=77pt,bblly=213pt,bburx=540pt,bbury=583pt,clip=}}
\vspace{0.1in}
\noindent
{\footnotesize {\bf FIG. 6(b).} Same as Fig.~6(a) for 
  ${\bf x} = (0.0223, 0.2)$, near the back wall.}
\label{fig_6b}
\end{figure}

\begin{figure}
\centerline{
\psfig{file=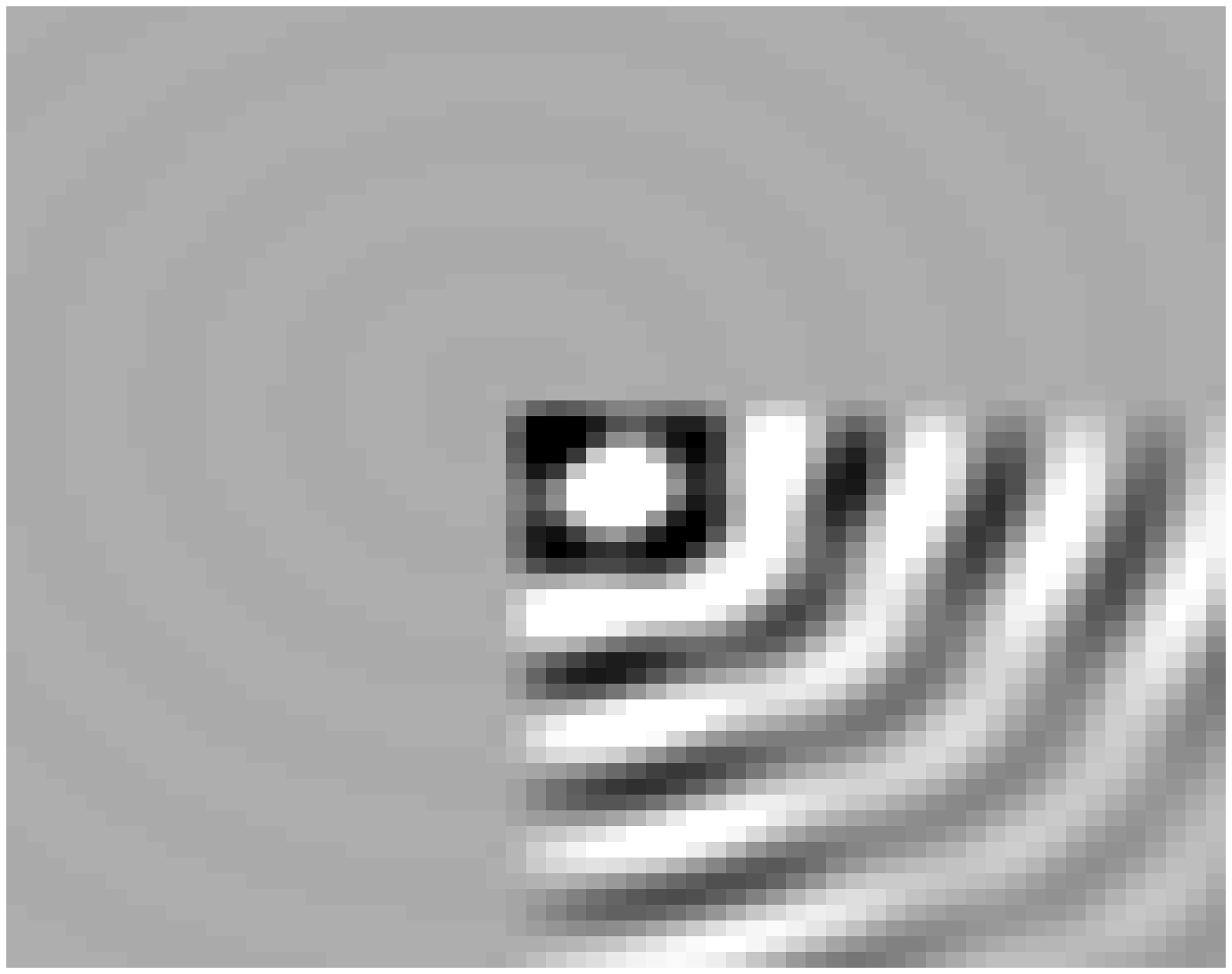,width=3in,bbllx=77pt,bblly=213pt,bburx=540pt,bbury=583pt,clip=}}
\vspace{0.1in}
\noindent
{\footnotesize {\bf FIG. 6(c).} Same as Fig.~6(a) for ${\bf x} = (0.0255, 0.0255)$, near the lower-left corner.}
\label{fig_6c}
\end{figure}

\begin{figure}
\centerline{
\psfig{file=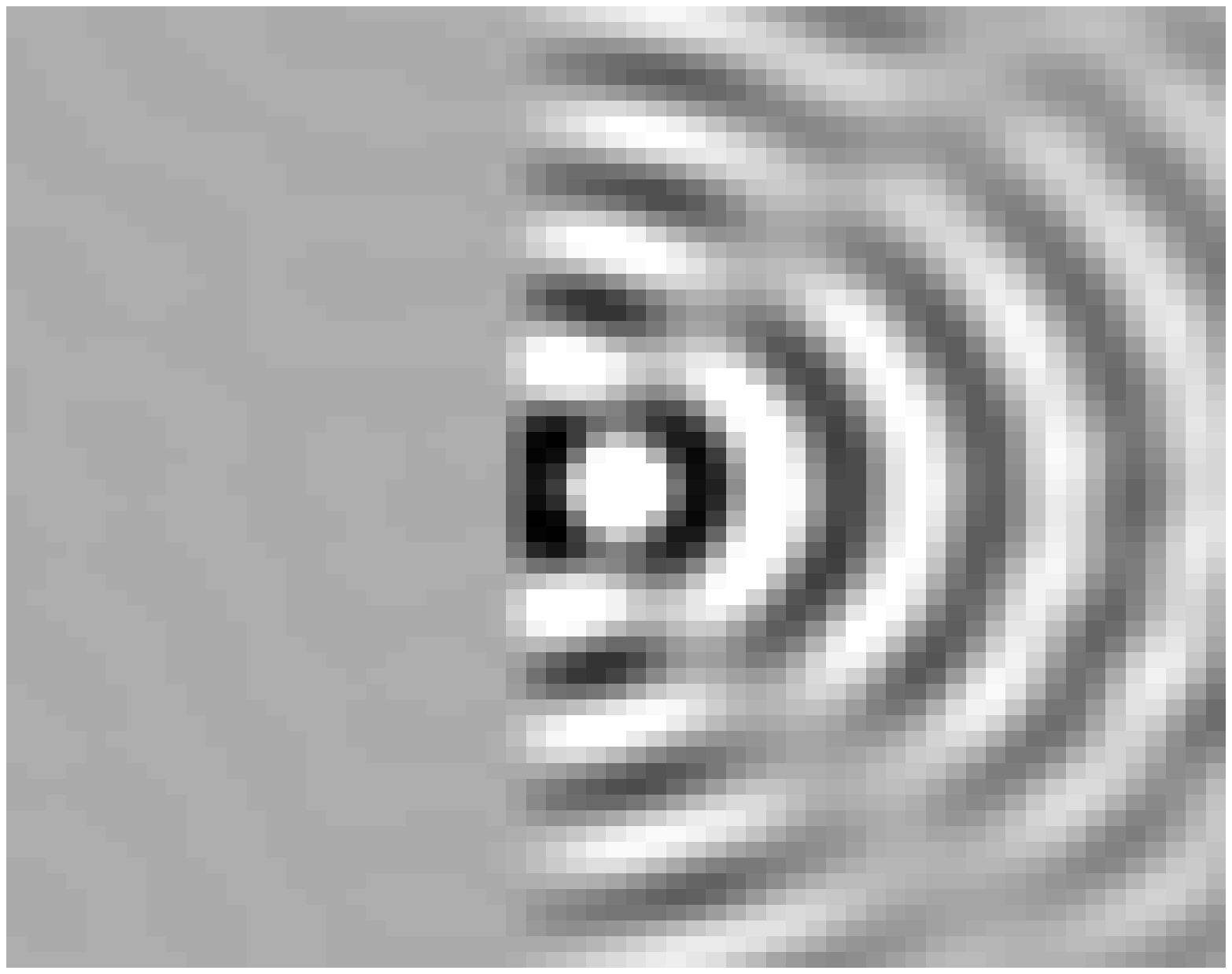,width=3in,bbllx=77pt,bblly=213pt,bburx=540pt,bbury=583pt,clip=}}
\vspace{0.1in}
\noindent
{\footnotesize {\bf FIG. 6(d).} Same as Fig.~6(a) for 
   ${\bf x} = (0.2, 0.0255)$, near the bottom side of the stadium.}
\label{fig_6d}
\end{figure}

\begin{figure}
\centerline{
\psfig{file=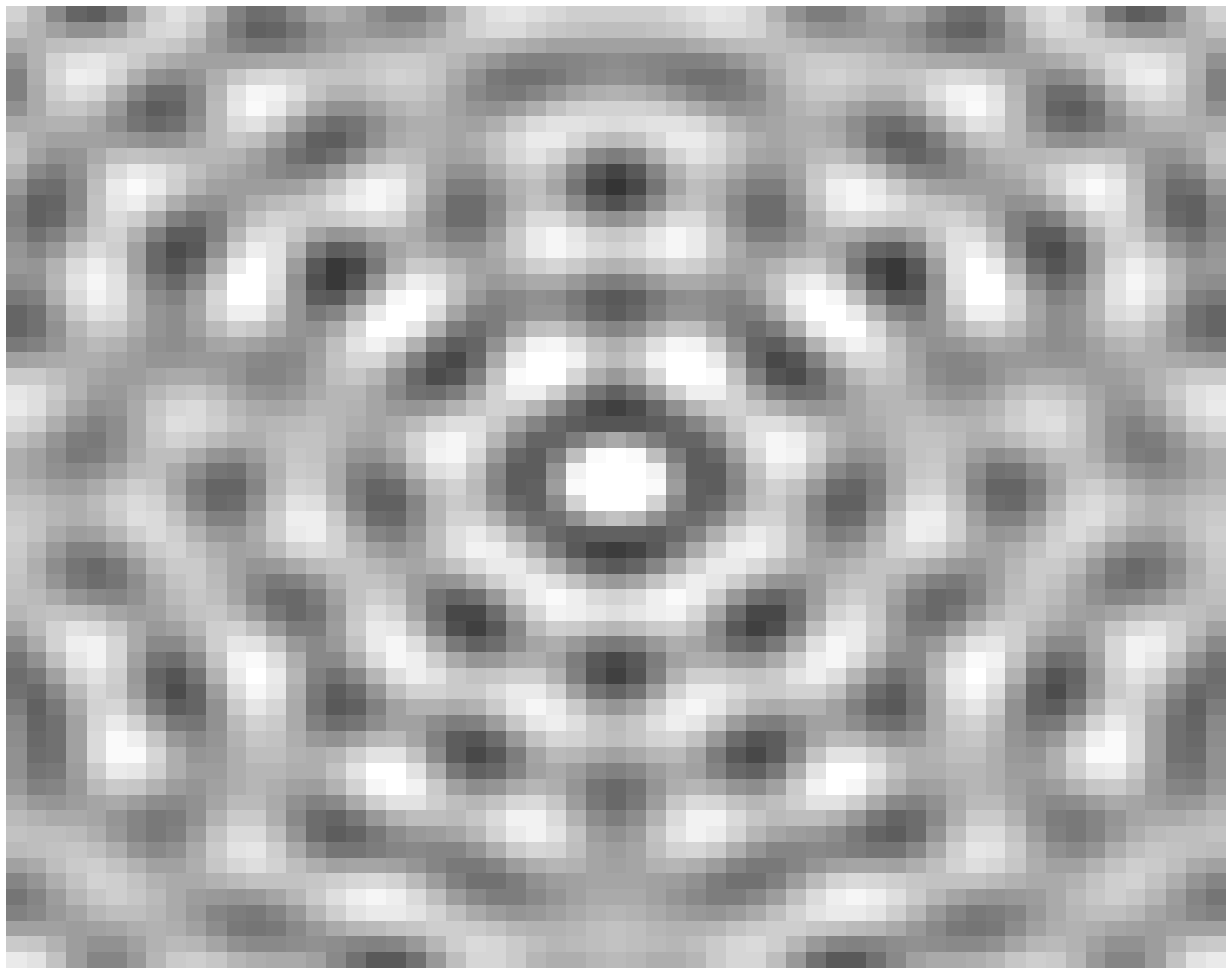,width=3in,bbllx=77pt,bblly=213pt,bburx=540pt,bbury=583pt,clip=}}
\vspace{0.1in}
\noindent
{\footnotesize {\bf FIG. 7(a).} 
  Theoretical correlation function for ${\bf x} = (0.3, 0.3)$ (same grid as
  in Fig.~6(a)). }
\label{fig_7a}
\end{figure}

\begin{figure}
\centerline{
\psfig{file=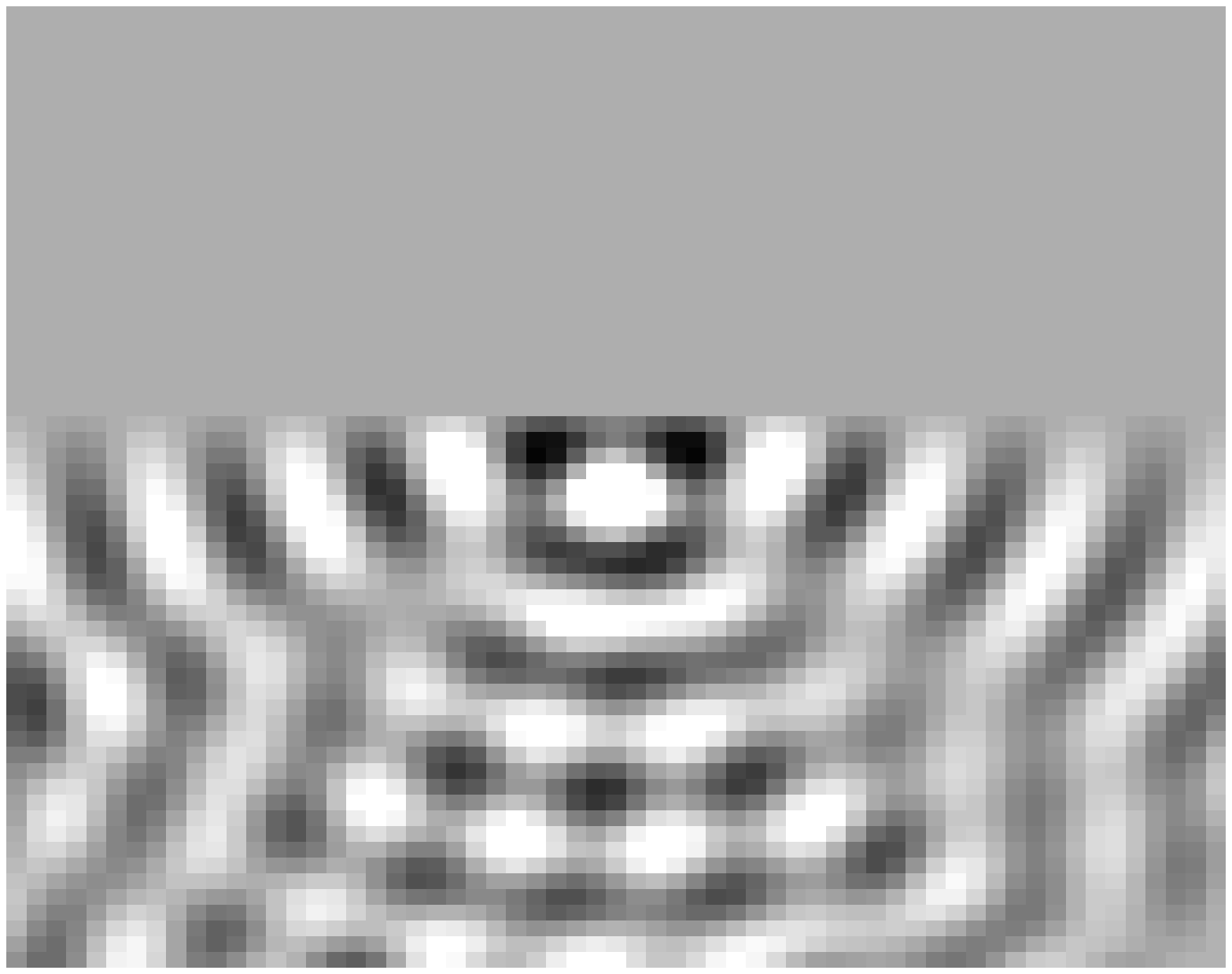,width=3in,bbllx=77pt,bblly=213pt,bburx=540pt,bbury=583pt,clip=}}
\vspace{0.1in}
\noindent
{\footnotesize {\bf FIG. 7(b).} Same as Fig.~7(a) for ${\bf x} = (0.0223, 0.2)$ (same grid as Fig.~6(b)).}
\label{fig_7b}
\end{figure}

\begin{figure}
\centerline{
\psfig{file=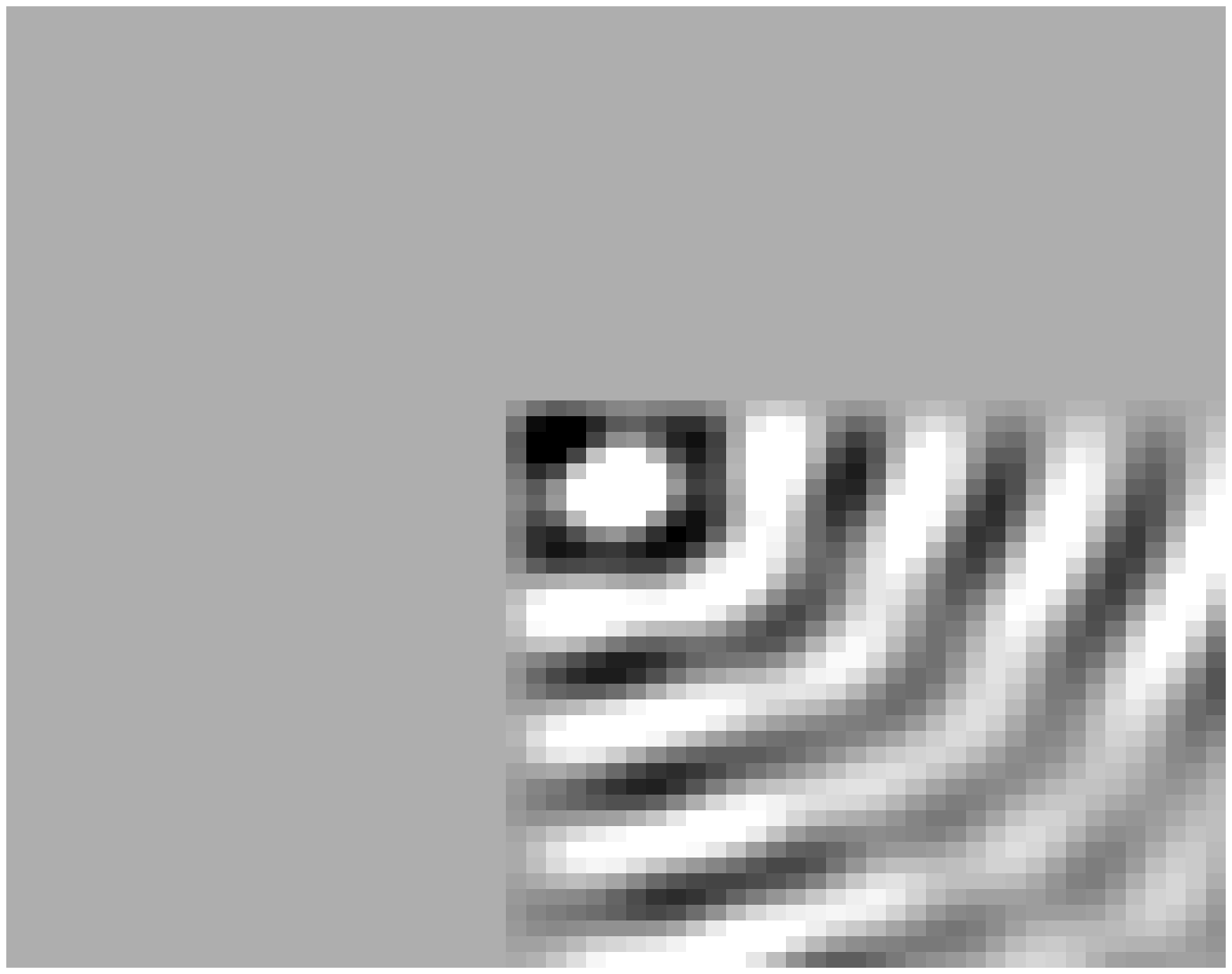,width=3in,bbllx=77pt,bblly=213pt,bburx=540pt,bbury=583pt,clip=}}
\vspace{0.1in}
\noindent
{\footnotesize {\bf FIG. 7(c).} Same as Fig.~7(a) for
 ${\bf x} = (0.0255, 0.0255)$ (same grid as Fig.~6(c)).}
\label{fig_7c}
\end{figure}

\begin{figure}
\centerline{
\psfig{file=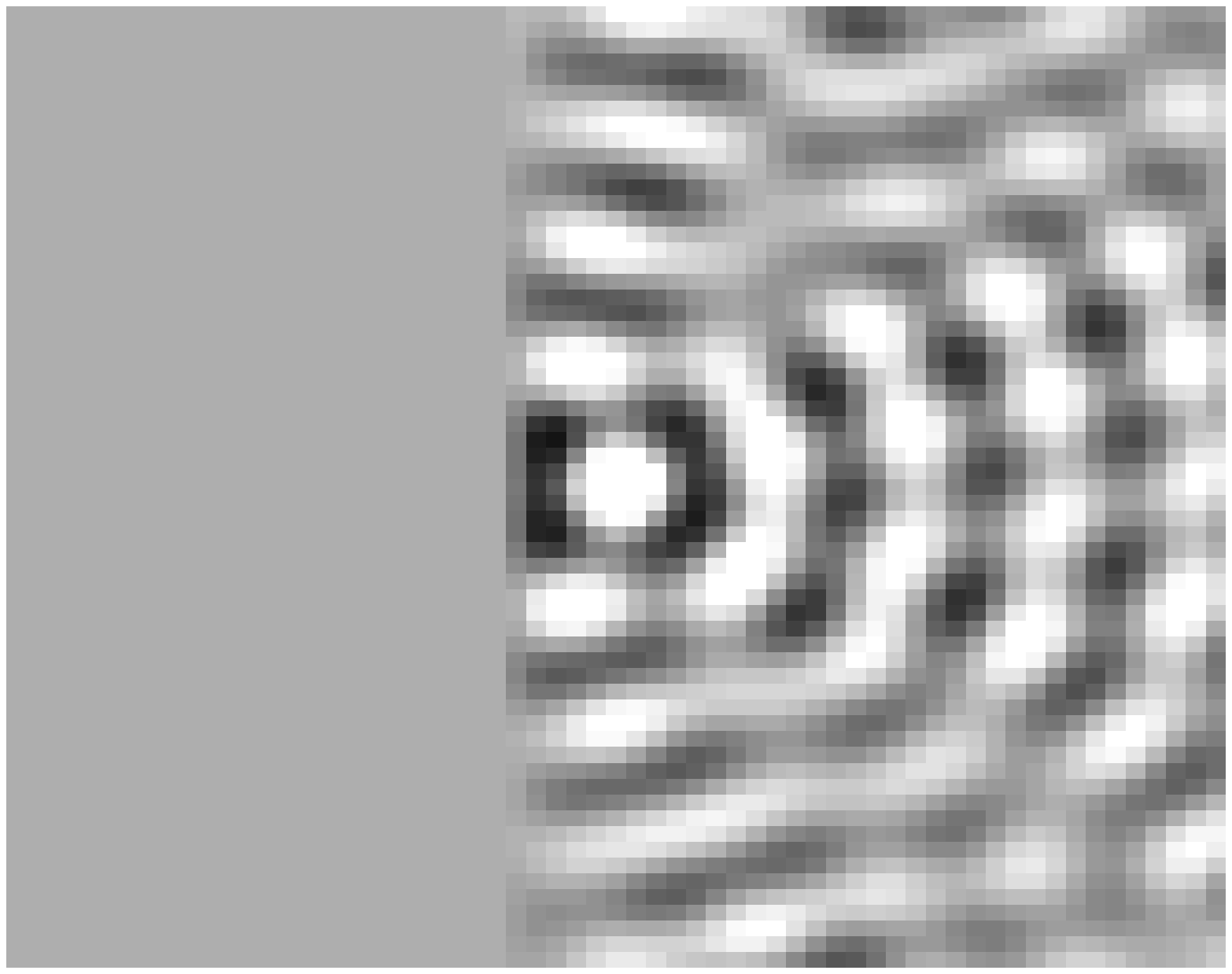,width=3in,bbllx=77pt,bblly=213pt,bburx=540pt,bbury=583pt,clip=}}
\vspace{0.1in}
\noindent
{\footnotesize {\bf FIG. 7(d).} Same as Fig.~7(a) for 
   ${\bf x} = (0.2, 0.0255)$ (same grid as Fig.~6(d)).}
\label{fig_7d}
\end{figure}

\end{document}